\documentclass{article}
\usepackage{graphicx}
\usepackage{subfigure}
\usepackage{url}
\usepackage{float}
\usepackage{multirow}
\usepackage{array}
\usepackage{hyperref}

\makeatletter
    \newcommand\figcaption{\def\@captype{figure}\caption}
    \newcommand\tabcaption{\def\@captype{table}\caption}
\makeatother

\begin{document}

\title{Characterization of Pentagons Determined by Two X-rays}
\author{Ming-Zhe Chen 
\\Qwest Communication Inc.
\\ 5751 Sells Mill Dr 
        \\ Dublin, OH 43017, USA}
\maketitle

\begin{abstract}
This paper contains some results of pentagons which can be 
determined by two X-rays. The results reveal this problem is more 
complicated.
\end{abstract}

\section{Introduction}
In computerized tomography, the structure of a planar convex body can
be determined by certain sets of four X-rays \cite{gardner}. But no 
accurate image, in general, can be determined by two X-rays. This "unique 
problem" can be back to the work of Lorentz \cite{lorentz}. A lot of 
works continued after that \cite{fishburn} \cite{gardner1} \cite{gardner2} 
 \cite{kuba} \cite{giering}, etc. The problem arises that in what conditions
or what is the character, the planar convex body can be determined by two X-rays.
Giering \cite{giering} gave partial results about triangles and quadrilaterals.
This paper gives some results of pentagons.
In the following, we give the triangle as an example to demonstrate the issue.
(For details about the Steiner symmetral of the triangle, see \cite{giering})
Figure 1 shows a Steiner symmetral of the triangle $A_{2}B_{1}C_{3}$ in the horizontal
direction; also it shows another Steiner symmetral of the triangle in the 
vertical direction.
 
\begin{figure}[H]
\begin{center}
\includegraphics[width = .5\textwidth, height = !]{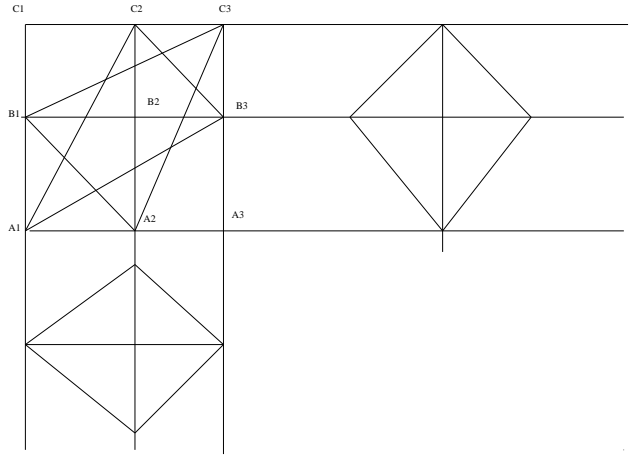}
\caption{An Example of Steiner Symmetrals of Triangles}\label{fig:mingzhechen1}
\end{center}
\end{figure}

In general, the triangle $A_{2}B_{1}C_{3}$ and the triangle
$A_{1}B_{3}C_{2}$ should have different Steiner symmetrals. But in some cases, such as,
if the chord of the triangle $A_{2}B_{1}C_{3}$ in the direction $B_{1}B_{2}B_{3}$ is
equal to the chord of the triangle $A_{1}B_{3}C_{2}$ in the direction $B_{1}B_{2}B_{3}$
and  the chord of the triangle $A_{2}B_{1}C_{3}$ in the direction $C_{2}B_{2}A_{2}$ is
equal to the chord of the triangle $A_{1}B_{3}C_{2}$ in the direction $C_{2}B_{2}A_{2}$
("equal chord" condition),
then the two different triangles $A_{2}B_{1}C_{3}$ and $A_{1}B_{3}C_{2}$ will have the same
 Steiner symmetrals; in other words, in this case we cannot determine the two triangles
by the Steiner symmetrals.

\section{The Case of Pentagons}
Details on the classification of pentagon cases are very complicated;
it is even more complicated if we mix pentagons with quadrilaterals.
Here we only investigate a case of two pentagons, which still reveals
very complicated conditions.
Figure 2 shows two pentagons: $A_{1}D_{2}E_{3}C_{5}B_{4}$ and 
$B_{1}C_{2}E_{4}D_{5}A_{3}$, $A_{1}A_{5}E_{5}E_{1}$ is a unity square. 

\begin{figure}[H]
\begin{center}
\includegraphics[width = .5\textwidth, height = !]{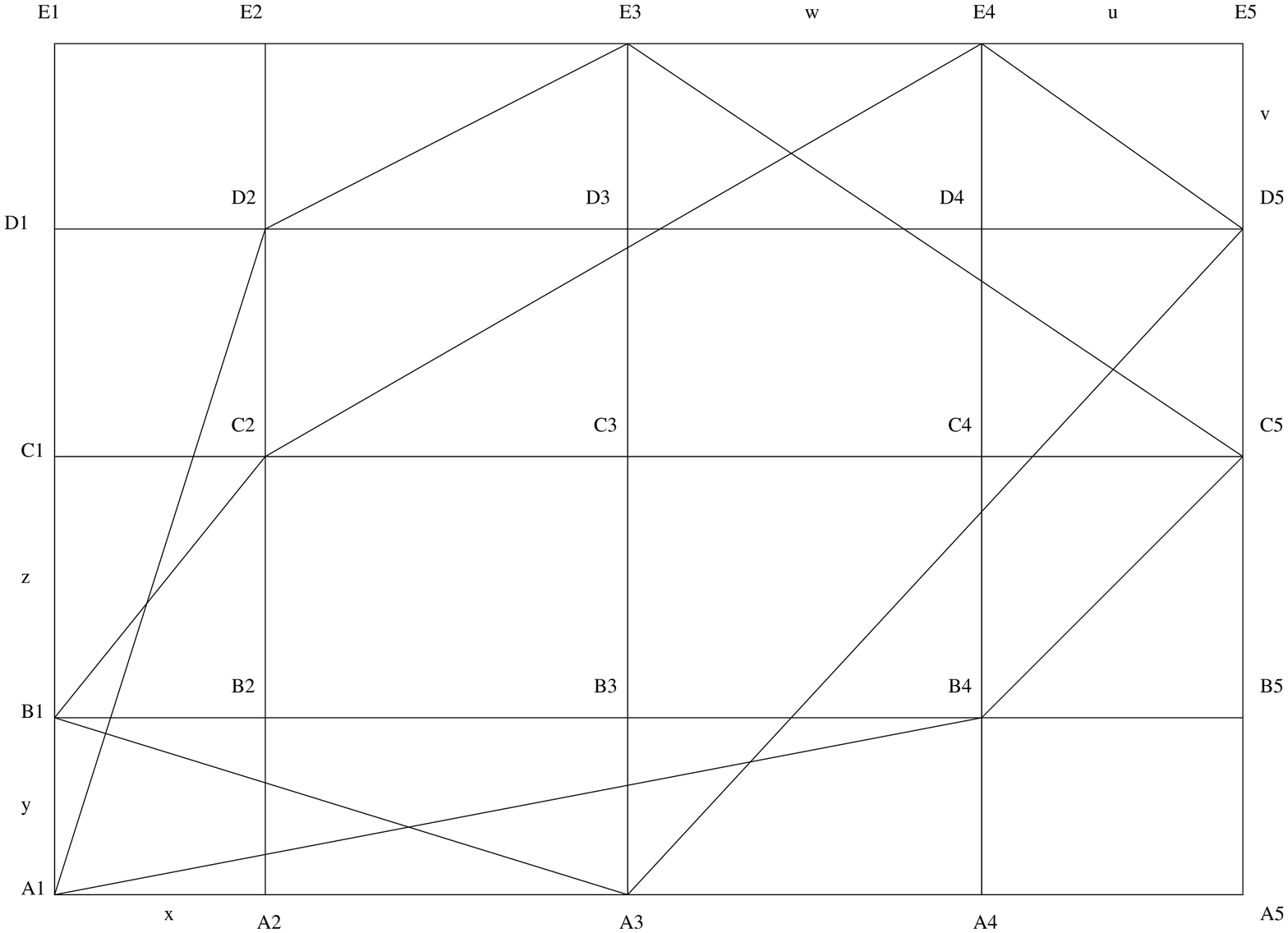}
\caption{An Example of Two Pentagons}\label{fig:mingzhechen2}
\end{center}
\end{figure}

From above "equal chord" condition, we can get the following related equations:
from the direction $B_{1}B_{5}$, we have
\[w - wv - xy - uy - wy = 0, \]
\[1 - v \ne 0   \]
from the direction $C_{1}C_{5}$, we have
\[u + w - x = 0, \]
\[1 - v \ne 0, \]
\[1 - v - y - z \ne 0   \]
from the direction $D_{1}D_{5}$, we have
\[v - 2uv - xv + 2u - 2yu - 2zu - wv + 1 + y + z + x - xy - xz + w - yw - zw = 0, \]
\[1 - y - z \ne 0   \]
from the direction $E_{2}A_{2}$, we have
\[1 - 2u - v + 2uv - 2xy - z + zw + 2zu + u^2 - u^2v + 2xyu - zwu - zu^2 \]
\[- w + wu + wv - uvw + xyw = 0, \]
\[1 - u \ne 0, \]
\[1 - w - u \ne 0   \]
from the direction $E_{3}A_{3}$, we have
\[-y + 2yu - yu^2 + xy - xyu - xyw + w - uw - zw + zwu = 0, \]
\[1 - u \ne 0, \]
\[1 - x - u \ne 0   \]
from the direction $E_{4}A_{4}$, we have
\[zw - yu - vw = 0, \]
\[u + w \ne 0   \]

After eliminating the variables z and w, we have
\[(y - 3yu + yu^2 - 3xy + 3xyu + u)v + (-y + 2yu + 3xy - 3xyu - 2x^2y^2 - u) = 0,   \]

\[(-4x + 7xu + 4x^2 - 2u^2 + u)v^2 + (-10xu - 2x^2 + 4u^2 - 2u - 2xy + 6xyu - 6x^2y)v \]
\[+ (3xu - 2u^2 + 2x^2y + 2x + 2xy + u + 2x^2 - 6xyu - 4x^2y^2) = 0,   \]

\[(4x - 8xu - u + 2u^2 + 4xu^2 - u^3)v^2 + (-6x + 12xu + 2u - 4u^2 - 6xu^2 + 2u^3 \]
\[+ 12x^2y - 12x^2yu - 2xyu + 2xyu^2)v \]
\[+ (2x - 4xu - 8x^2y - u + 2u^2 + 2xu^2 + 8x^2yu - u^3 + 2xyu - 2xyu^2 + 4x^3y^2) = 0   \]

When we need to reconstruct or determine a pentagon from Steiner symmetrals,
we can draw a pentagon as one in the Figure 2 using Steiner symmetrals (chords of
a pentagon to be reconstructed can be obtained by translating chords of Steiner symmetrals
 in the directions).
Then we can get the values of the variables u and v, and then we can use the values of u and v
and above three related equations to plot a curve of (x, y).
If the point $B_{2}$(x, y)(Figure 2) happens to be on the curve we plotted, it means the pentagon
we just reconstructed is not unique; there are a group of pentagons 
having the same Steiner symmetrals.

We can further eliminate the variables v from above three related equations.
After eliminating the variables v, we have
\[(-2y^2)u^6 + (5y^2 - 11xy^2 - 6xy^3)u^5  \]
\[+ (-4y^2 - 4xy - 10x^2y^2 - 8x^2y^3 - 2xy^2 + 26xy^3 - 4x^2y^4)u^4 \] 
\[+ (y^2 + 26x^2y^2 + 8xy - 13xy^2 - 32xy^3 + 6x^3y^2 + 2x^2y - 26x^3y^3 + 22x^2y^3 - 12x^3y^4 + 24x^2y^4)u^3 \]
\[+ (-16x^4y^3 + 8xy^2 + 14xy^3 + 24x^3y + 36x^4y^2 + 4x^2 - 34x^2y + 4xy - 2x - 94x^3y^2 + 52x^3y^3 \]
\[+ 26x^2y^2 - 22x^2y^3 + 56x^3y^4 - 20x^4y^4 - 44x^2y^4)u^2 \]
\[+ (36x^5y^3 - 8x^5y^4 - 32x^4y^3 - 2xy^3 - 60x^4y^2 - 24x^3y + 122x^3y^2 + 20x^2y - 4xy \]
\[ + 4xy^2 - 18x^3y^3
- 50x^2y^2 + 16x^2y^3 + 64x^4y^4 - 72x^3y^4 + 24x^2y^4)u \]
\[+ (16x^6y^4 - 36x^5y^3 + 60x^4y^3 + 20x^5y^4 + 36x^4y^2 + 16x^2y^2 - 42x^3y^2 - 2xy^2 \]
\[+ 8x^3y^3 - 36x^4y^3 
- 4x^2y^3 + 20x^3y^4 - 36x^4y^4 - 4x^2y^4) = 0,   \]

\[(-y)u^7 + (4y + 2xy - 2xy^2)u^6 + (-6y - 6x^2y - 2x - 6xy + 10xy^2 + 6x^2y^2)u^5 \]
\[+ (4y + 6x + 2xy - 16xy^2 + 28x^2y - 26x^2y^2 + 8x^3y^3)u^4 \]
\[+ (-y - 28x^2y - 6x + 8xy + 10xy^2 + 22x^2y^2 - 24x^4y^2 - 4x^2 - 20x^3y + 8x^4y^3 + 52x^3y^2 - 40x^3y^3)u^3 \] 
\[+ (4x^2y + 2x - 8xy - 20x^5y^3 - 2xy^2 + 44x^3y - 128x^3y^2 + 72x^4y^2 + 4x^2 
+ 10x^2y^2 - 16x^4y^3 + 60x^3y^3)u^2 \]
\[+ (2xy + 40x^5y^3 - 72x^4y^2 - 20x^3y - 16x^2y^2 + 88x^3y^2 - 28x^3y^3 + 8x^4y^3 + 2x^2y)u \]
\[+ (4x^3y^3 + 4x^2y^2 + 24x^4y^2 - 20x^3y^2 - 20x^5y^3) = 0   \]

We can further reduce above two equations to be one,
which is (x, y) curve, 
using the method of the resultants, but the result is too complicated to be 
written here. We just show the first term of the equation as follows:
\[16^7x^{42}y^{34} + \dots \]

\bibliography{mingzhechen}
\bibliographystyle{plain}
\end{document}